# Enhancement of the upper critical field of Nb$_3$Sn utilizing disorder introduced by ball milling the elements

L. D. Cooley, Y. F. Hu, and A. R Moodenbaugh

*Condensed Matter Physics and Materials Science Department, Brookhaven National Laboratory, Upton NY 11973*

Nb$_3$Sn was prepared by milling Nb and Sn powder mixtures followed by limited reactions to restrict disorder recovery. Although disorder reduced the superconducting critical temperature $T_c$, the concomitant electron scattering increased the upper critical field $\mu_0 H_{c2}$ to as high as 35 T at 0 K, as determined by the Werthamer-Helfand-Hohenberg equation. $H_{c2}$ was higher for longer milling times and lower annealing temperatures. Substitution of 2% Ti for Nb did not appreciably enhance $H_{c2}$, suggesting that alloying mitigates the benefits of disorder. Since alloyed Nb$_3$Sn wires have $\mu_0 H_{c2}(0) \approx 29$ T, wires based on heavily milled powders could extend the field range for applications if they can be made with high current density.

Practically all high-field superconducting magnets operating above ~8 T at 4.2 K use Nb$_3$Sn wires. This includes laboratory solenoids, large solenoids for fusion, dipoles and quadrupoles for particle accelerators, and very high field magnets for nuclear magnetic resonance. The operating envelope for Nb$_3$Sn, presently ~8 < $\mu_0 H$ < ~21 T at 4.2 K (up to ~23 T at 2.0 K), is determined by the temperature dependence of the upper critical field $H_{c2}(T)$ and the field $H$ and temperature $T$ dependence of the critical current density $J_c$. While $J_c$ depends strongly on the development of a homogeneous fine-grained microstructure through control of the Nb$_3$Sn formation reaction, $H_{c2}$ is determined by the physics of the superconducting state itself.

Very recently, Godeke *et al.* examined the factors that affect the $H_{c2}(T)$ curve in a range of commercial and experimental Nb$_3$Sn composite wires.[1] Generally, their results fall into two classes: 1) pure Nb$_3$Sn wires which exhibit $T_c \approx 18.0$ K, $\mu_0 H_{c2}' \equiv \mu_0 dH_{c2}/dT|_{T_c}$ of about -2.0 T K$^{-1}$, and an extrapolated $\mu_0 H_{c2}(0) \approx 26$ T; and 2) alloyed wires with 1 to 2% Ti or Ta substituted for Nb which exhibit $T_c \approx 17.5$ K, $\mu_0 H_{c2}' \approx -2.3$ T K$^{-1}$, and $\mu_0 H_{c2}(0) \approx 29$ T. These wires' behaviors were described very well by theoretical descriptions developed during the 1960s, in particular the Werthamer, Helfand, and Hohenberg (WHH)[2] equation

$$\mu_0 H_{c2}(0) = 0.69\, \mu_0\, H_{c2}'\, T_c. \qquad (1)$$

Since $H_{c2}'$ is directly proportional to the normal-state electrical resistivity $\rho_N$, alloying the Nb$_3$Sn phase with Ti and/or Ta adds electron scattering and improves $H_{c2}$ at low temperatures. Alloying also changes the electronic density of states and the electron-phonon coupling, since adding more than about 2% of Ti or Ta results in a drop of $T_c$ that more than counteracts the increase in $H_{c2}'$. Hence, the maximum observed enhancement is 2-3 T at 4.2 K.[3]

In this Letter, we report that greater improvements in $\mu_0 H_{c2}(0)$ can be realized by introducing disorder instead of by alloying. Niu and Hampshire showed recently that high-energy ball milling of PbMo$_6$S$_8$ superconductors, followed by controlled annealing in a hot isostatic press, resulted in a tripling of $H_{c2}'$ at the expense of a modest reduction of $T_c$ from 15 to 12 K.[4] The $\mu_0 H_{c2}(0)$ values estimated by the WHH formula increased from 45 to 110 T. Based on this result, we explored whether a similar behavior could be induced in Nb$_3$Sn.

Past experiments have shown that the superconducting properties of Nb$_3$Sn are optimized for compositions found at the tin-rich (stoichiometric) edge of the phase field.[5] Even for the stoichiometric composition, the superconducting properties are very sensitive to atomic site disorder introduced by irradiation or strain,[6] suggesting that the decrease in $T_c$ due to disorder might more than offset the benefit from the increase of $H_{c2}'$. Several researchers studied the development of disorder as Nb$_3$Sn powders were milled by comparing $T_c$ and x-ray diffraction patterns.[7,8,9] These studies showed that the disorder was indeed very potent at reducing $T_c$, and that the recovery of $T_c$ required annealing at higher temperatures and for longer times (i.e. 700 to 1000 °C for many hours) than typical Nb$_3$Sn reactions. The disordered material was thought to be akin to Nb$_3$Sn damaged by radiation. Kim and Koch also showed that Nb$_3$Sn could be formed directly by mechanical alloying of Nb and Sn powders.[10] However, measurements of $\rho_N$ or $H_{c2}$ were not reported.

In the present experiment, high-energy ball milling was used to blend Nb (99.9%, -325 mesh, Oremet Wah Chang) and Sn (99.99%, -100 mesh, Ventron) powders into a nanoscale mixture. Nb$_3$Sn was then formed during an anneal. Binary mixtures of 0.75Nb + 0.25Sn were milled for 2 and 6 hours (sample series beginning B2 and B6, respectively). Ternary samples were made with 2% Ti (99.9%, -325 mesh, Cerac) powder substituted, i.e. 0.73Nb + 0.02Ti + 0.25Sn. Tungsten carbide vials and balls were used with a 4:1 mass ratio of ball to total powder, in a Spex model 8000M dual ball mill with forced air cooling. Agglomeration and caking of the products accounted for typically < 10% of the total milled product mass. The milled mixtures were then pressed into 9 mm diameter pellets using a piston press with 20 to 25 kN force. These were placed into a high-vacuum furnace and reacted using one of four schedules: (A) 600 °C for 24 h, (B) 650 °C for 24 h, (C) 750 °C for 10 h, or (D) 950 °C for 1 h. Thus sample B2A denotes a binary mixture milled for 2 hours,

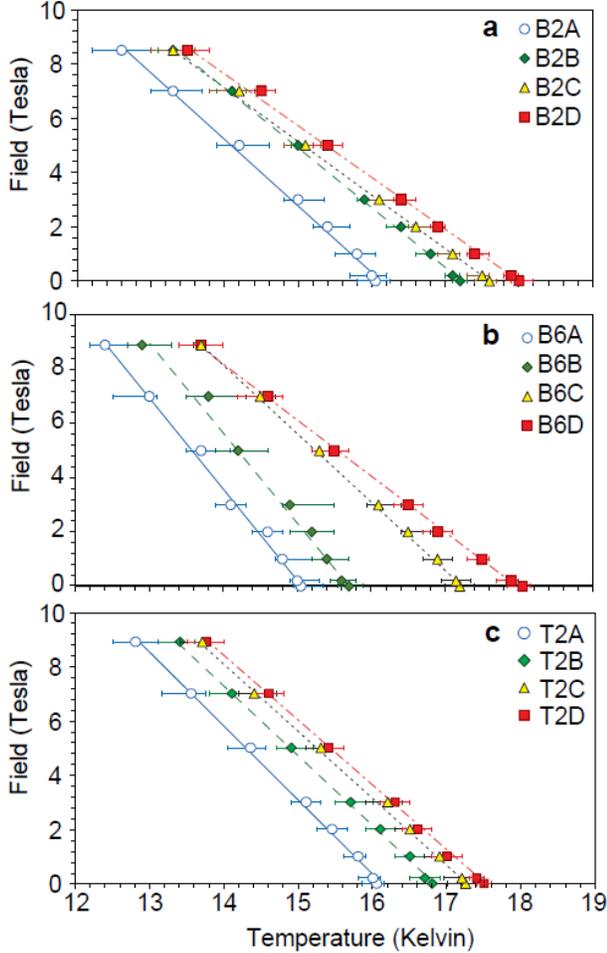

Figure 1. $H_{c2}(T)$ as determined from resistive measurements of the superconducting transition. Plots (a) and (b) show data for samples prepared from binary powder milled 2 and 6 hours, respectively, while plot (c) shows data for samples prepared from ternary powder milled 2 hours. Data points and error bars denote 50% and the interval from 10% to 90% of the normal state resistivity, respectively. Lines are least-squares fits to the data

then reacted at 600 °C for 24 hours. The resulting products were ~65% dense, based on pellet mass and volume.

X-ray diffraction patterns (Cu Kα radiation) of the as-milled powders showed that the principal peaks of Nb were significantly broadened, while the peaks for Sn were absent. This finding is different from that of Kim and Koch,[10] who observed peaks for both elements for comparable milling times, but using steel balls. Following the observation reported in ref. 8, it appears that Sn is more quickly converted to an amorphous phase when processed with the heavier WC balls. The Williamson-Hall formalism[11] was used to analyze the diffraction peak broadening to estimate particle size, which was ~14 nm after 2 h milling and ~4 nm after 6 h. Weak peaks of both $Nb_3Sn$ and $Nb_6Sn_5$ began to appear after 4 hours of milling, consistent with the study of Kim and Koch.[10] Powder diffraction after the anneal indicated conversion into $Nb_3Sn$, sometimes with additional peaks indicating <5% $Nb_6Sn_5$ as a trace phase. Some broadening remained for the more heavily milled and less well annealed samples, consistent with their having higher disorder.

For superconducting property characterization, bars with ~ 1 mm² cross section and 5 mm length that were cut from the reacted pellets. Silver paint was used to make connections for four-probe measurements, with contact resistances from 0.1 to 10 ohms. The electrical resistivity $\rho(H,T)$ was measured with 0.1 A constant current applied perpendicular to a fixed background field of $0 < H < 8.9$ T, while sweeping the temperature over $10 < T < 20$ K. During the initial cooling from room temperature, $\rho(0,T)$ decreased with decreasing $T$ for all samples.

Fig. 1 summarizes the $\rho(H,T)$ measurements, showing data points corresponding to 50% of $\rho_N$ and error bars representing the span from 10% to 90% of $\rho_N$. In each plot, the lines represent least-squares fit to the data points. All three plots show that the slope of $H_{c2}(T)$ decreases with more intensive annealing for fixed milling time and composition. Also, by comparing figs. 1(a) and 1(b), one observes that, for fixed annealing conditions, $H_{c2}'$ for each B6 sample is greater than that of the corresponding B2 sample. Both of these trends are consistent with the earlier observation by Niu and Hampshire,[4] that electron scattering is increased by disorder for longer milling time and lower intensity annealing.

The trends from Fig. 1 are summarized in Table I. From these data, values of $\mu_0 H_{c2}(0)$ were calculated using Eq. (1). The most important results are the ~35 T values of $\mu_0 H_{c2}(0)$ for samples B6A and B6B. These are about 6 T higher than for Ti-alloyed $Nb_3Sn$ (~29 T, without strain due to contact with bronze),[3] and 8 to 10 T higher than the well-annealed binary samples. Samples B6A and B6B have $\mu_0 H_{c2}(0)$ values that are also ~5 T higher than the best values reported by Godeke et al. for state-of-the-art wires.[1] This demonstrates that disorder can be used to enhance $H_{c2}$ above the level currently being obtained by other routes. Table I shows that increasing the amount of retained disorder results in a dramatic increase in $\mu_0 H_{c2}'$, to a maximum (absolute value) here of about 3.3 T K⁻¹. A moderate drop in $T_c$ accompanies this trend; $T_c = 15.1$ K for the most strongly disordered sample.

Table I also shows that the estimated $\rho_N$ at 20 K for our samples increases in concert with $H_{c2}'$. Since the porosity and existence of impurity phases make the cross-section through which the current passes uncertain, we scaled our measured values to make the well annealed samples B2D, B6D, and T2D have $\rho_N$ values consistent with known values[3] for binary and ternary $Nb_3Sn$ with similar $T_c$ and $\mu_0 H_{c2}'$. Since this scaling factor was approximately 0.3 in all cases, we subsequently multiplied all of the measured $\rho_N$ values by 0.3 to give the values reported in Table I. Note that the maximum $\rho_N$ estimated for $T = 20$ K (300 μΩ-cm) is almost 5 times that obtained in alloyed $Nb_3Sn$ wires, and is well above the ~140 μΩ-cm range where the electron mean free path approaches the interatomic spacing.[12] Therefore, the apparent magnitude of $\rho_N$ should be viewed cautiously, as suggested recently for $MgB_2$ superconductors under similar circumstances.[13]

Table I. Properties of reacted samples.

| Sample ID | $T_c$ (K) | $\rho_N$ ($\mu\Omega$-cm) | $\mu_0 H_{c2}'$ (T K$^{-1}$) | $\mu_0 H_{c2}(0)$ (T) |
|---|---|---|---|---|
| B2A | 15.7 | 140 | -2.45 | 27.3 |
| B2B | 16.7 | 120 | -2.20 | 26.6 |
| B2C | 17.6 | 35 | -2.08 | 26.3 |
| B2D | 18.0 | 26 | -1.93 | 24.0 |
| B6A | 15.1 | 300 | -3.34 | 35.7 |
| B6B | 15.8 | 190 | -3.29 | 34.8 |
| B6C | 17.2 | 170 | -2.53 | 30.1 |
| B6D | 18.0 | 39 | -2.07 | 25.6 |
| T2A | 16.1 | 75 | -2.71 | 30.2 |
| T2B | 16.8 | 56 | -2.59 | 30.0 |
| T2C | 17.2 | 50 | -2.45 | 29.1 |
| T2D | 17.5 | 48 | -2.40 | 29.0 |
| $Nb_{0.76}Ti_{0.02}Sn_{0.22}$ (ref. 3) | 17.6 | 45 | -2.38 | 28.9 |

The ternary samples display a much weaker response to disorder than the binary samples for 2 hrs. milling time and identical annealing [fig. 1(a) vs. 1(c)], with $\mu_0 H_{c2}(0)$ values falling within a 1 T range in Table I. We speculate that two mechanisms are nearly balanced in the T2 samples. In the well-annealed sample T2D, Ti substitutes for Nb, causing electron scattering but leaving Nb atom chains (which are important for superconductivity) relatively intact.[3] Disordered samples T2A-C and B2A-C, on the other hand, should contain interstitial atoms and anti-site disorder, similar to the defects produced by irradiation. This both adds scattering and reduces $T_c$. Since the values of $T_c$ and $\rho_N$ in Table I indicate more disorder in B2A and B2B than in T2A and T2B, Ti must somehow increase disorder recovery or impede its formation. This negates the additional electron scattering added by disorder. Another possibility is that Ti causes defects to cluster, producing pockets in which electron scattering is dominated by disorder surrounded by a continuous region where electron scattering is dominated by substitutional alloying.

The present result is different from earlier reports on irradiated samples. Although the lack of full sample connectivity prevents us from accurately determining $\rho_N$, the relative changes of $\rho_N$ with respect to $T_c$ are much stronger in our samples than in past work on irradiated Nb$_3$Sn films. For example, sample B6A retains a $T_c$ = 15.1 K even though it is extensively disordered. In contrast, films that were irradiated by a sufficient fluence of alpha particles to reduce $T_c$ to 4 K exhibited a $\rho_N$ of 130 $\mu\Omega$-cm.[12] The maximum absolute value of $\mu_0 H_{c2}'$ reported for those irradiated films was less than 3.0 T K$^{-1}$,[12] significantly inferior to the absolute values of ~3.3 T K$^{-1}$ reported here. A possible explanation of these differences is that defects formed by irradiation, upon annealing, form clusters, which then frees up short-circuit pathways through adjacent material with less damage. In contrast, the ball-milled samples may be disordered uniformly and recover at about the same rate everywhere in the sample.

In summary, Nb$_3$Sn samples prepared using high-energy ball milling exhibit upper critical fields superior to those obtained in state-of-the-art wires. Some of the disorder incorporated into a mixture of Nb and Sn powders is retained by limiting the reaction anneal. Slopes of $H_{c2}(T)$ near $T_c$ are much higher than those reported previously, either for alloyed or for irradiated specimens. When these slopes are combined with the modest reduction of $T_c$, the Werthamer-Helfand-Hohenberg formalism predicts $\mu_0 H_{c2}(0)$ values as high as 35 T. The scattering mechanism is likely to be more like that produced by irradiation than that produced by alloying, based on the lack of significant changes when Ti was alloyed into ball-milled samples. The ~5 T increase in $\mu_0 H_{c2}(0)$ over that of wires could be a significant boost for high-field magnet applications.

This work was supported by the US Department of Energy under contract number DE-AC02-98CH10886. We would like to thank A. Ghosh, D. Welch, M. Suenaga, L. Snead, H. Wiesmann, M. Strongin, D. Larbalestier, M. Jewell, and B. Senkowicz for helpful discussions and analyses.